%% file: VolumeOperator.short.arxiv.tex
\documentclass[aps,nofootinbib,pra,preprint,showpacs,
longbibliography,
]{revtex4-1}

\usepackage[T1]{fontenc}
\usepackage[utf8x]{inputenc}
\usepackage{amsmath}
\usepackage{verbatim}

\usepackage{array}

\usepackage{color}
\usepackage[all]{xy}

\usepackage{tikz}

\usetikzlibrary{arrows}
\usetikzlibrary{shapes}
\usetikzlibrary{automata}
\tikzstyle{model_node}=[circle,draw=black,fill=tuggreen!50,minimum size=50pt]
\tikzstyle{tool}=[draw,double,rounded corners,inner sep=10pt]
\tikzstyle{aut_node}=[circle,draw=black,fill=black,inner sep=0pt,minimum size=7pt,font=\footnotesize]

\tikzset{
	MyPersp/.style={scale=0.8,x={(-0.8cm,-0.4cm)},y={(0.8cm,-0.4cm)},
    z={(0cm,1.0cm)}},
PerspFlatFig1/.style={scale=8.0,x={(0cm,0cm)},y={(1cm,0cm)},
    z={(0cm,1cm)}}, 
Persp3DFig2/.style={scale=10.0,x={(-0.8cm,-0.2cm)},y={(0.8cm,-0.2cm)},
    z={(0cm,1.0cm)}},
MyPoints/.style={fill=white,draw=black,thick}
		}
		
\pgfmathdeclarefunction{xTriang}{3}{%   	 xTriag(a,b,x)
  \pgfmathparse{0.5*(#2*#2+#3*#3-#1*#1)/#3}%
}

\pgfmathdeclarefunction{yTriang}{3}{%   	 yTriag(a,b,x)
\pgfmathparse{sqrt((1/4.0)*(#2+#1+#3)/(#3)*(#2+#1-#3)*(#2+#3-#1)*(#3-#2+#1)/(#3)} %
}

\makeatletter

\usepackage{multirow}

\usepackage{hyperref}

\makeatother

\begin{document}

\title{Hamiltonian dynamics of a quantum of space:\\ hidden symmetries and spectrum of the volume operator, \\ and discrete orthogonal polynomials}
%\date{}

\author{Vincenzo Aquilanti}
\affiliation{Dipartimento di Chimica, Università di Perugia, 06123 Perugia, Italy \\ and  Istituto di Metodologie Inorganiche e Plasmi, C.N.R., 00016 Roma, Italy }

\author{Dimitri Marinelli}
\affiliation{Dipartimento di Fisica, Universit\`a degli Studi
di Pavia, via A. Bassi 6, 27100 Pavia, Italy and INFN, Sezione di Pavia}

\author{Annalisa Marzuoli}
\affiliation{Dipartimento di Matematica `F. Casorati', Universit\`a degli Studi
di Pavia, \\ via Ferrata 1, 27100 Pavia, Italy  and INFN, Sezione di Pavia}

\begin{abstract}
The action of the  quantum mechanical volume operator, introduced in connection with a symmetric representation of the three-body problem and recently recognized to play a fundamental role in discretized quantum gravity models, can be given as a second order difference equation which,
by a complex phase change,  we turn into a 
discrete Schrödinger-like equation. The  introduction of discrete potential--like functions reveals the surprising  crucial role here of hidden symmetries, first discovered by  Regge  for the  quantum mechanical $6j$ symbols; 
insight is provided into the underlying geometric features.
The spectrum and wavefunctions of the volume operator are discussed
from the viewpoint of the Hamiltonian evolution of an elementary ``quantum of space'', and a transparent asymptotic picture emerges of the semiclassical and classical regimes. The definition of coordinates adapted to Regge symmetry is exploited for the
construction  of  a novel set of  discrete orthogonal polynomials, characterizing the oscillatory components of torsion-like modes.
\end{abstract}
\pacs{31.15-p; 03.65.Sq;  02.30.Gp; 04.60.Nc }
\maketitle
%
%
%\vfill
%\newpage
%
\section{Introduction}
Extension of  familiar angular momentum theory to describe quantum dynamics as a function of discrete variables is required to cope 
\textit{e.g.}
\emph{(i)}-with structure and reactivity in molecular, atomic and nuclear physics \cite{ springerlink:10.1007/s00214-009-0519-y,Odake2011Discrete}, 
\emph{(ii)}-with use in quantum chemistry of elliptic coordinates and orbitals \cite{PhysRevLett.80.3209,QUA:QUA10497}, %,aquilanti:4066}, 
\emph{(iii)}-with spin network approaches to quantum gravity \cite{Rovelli2007QG,QuantumTriangulations2012}. These approaches exploit progresses in understanding	solvability of quantum systems, such as provided by dynamical symmetry algebras \cite{ Granovskii19921,Odake2011Discrete}.

An elementary spin network  picture is schematized in Fig. \ref{fig:Quadrilateral}: alternatively  to the traditional sequential coupling of angular momenta, the volume operator $K=\mathbf{J}_1\cdot \mathbf{J}_2\times\mathbf{J}_3$, first defined in \cite{MR0194029}, acts democratically on  vectors $\mathbf{J}_1,\,\mathbf{J}_2$ and $\mathbf{J}_3$ plus a fourth one,  $\mathbf{J}_4$, which closes 
a (not necessarily planar) quadrilateral vector diagram ${\bf J_1 +J_2 + J_3 +J_4}=0$. 
Matrix elements of $K$ were computed in \cite{JeanMarcLevyLeblond1965Symmetrical} to provide a Hermitian representation, whose features have been studied by many 
\cite{
Rovelli1994Discreteness,*ErrataRovelliSmolin1995,*DePietri1996Geometry,*Ashtekar1997Quantum,
*Loll1995Volume,
Brunnemann2004Simplification,*Brunnemann2010Properties,*Brunnemann2010Properties1,
Meissner:2005mx} (see also \cite{HalBianchi2011,*Bianchi2012BohrSommerfeld} for an approach based on Bohr-Sommerfeld quantization). Carbone et al. \cite{Carbone2002Quantum} gave a geometrically based (Ponzano-Regge \cite{Ponzano1968Semiclassical}, Schulten-Gordon \cite{Schulten1975Semiclassical}) WKB asymptotics. 

In this work, by a suitable complex change of phase, we transform the imaginary antisymmetric representation into a  real, time-independent Schrödinger  equation which governs  the Hamiltonian dynamics as a function of a discrete variable denoted $\ell$. The Hilbert space spanned by the eigenfunctions of the volume operator \cite{Aquilanti2012Semiclassical} is constructed combinatorially and geometrically, applying polygonal relationships to the two quadrilateral vector diagrams in Fig. \ref{fig:Quadrilateral}, which are ``conjugated'' by a hidden symmetry discovered by Regge \cite{Regge1959Simmetry}.
We analyze the consequences of this elusive and other symmetries, providing a perspective geometrical view, helpful
for both the characterization of molecular spectra and torsion-like modes, and also for the extraction of the polynomial components of rotovibrational wavefunctions.

%\emph{Discrete Schrödinger equation and Regge symmetry -} 
\section{Discrete Schrödinger equation and Regge symmetry} 
Eigenvalues $k$ and  eigenfunctions $\Psi^{(k)}_{\ell}$  of the volume operator are most simply obtained through the three--terms recursion relationship 
first introduced in \cite{JeanMarcLevyLeblond1965Symmetrical} and analyzed in  \cite{Carbone2002Quantum}, where  analytical expressions of eigenvalues $k$ and eigenvectors  $\Psi^{(k)}_{\ell}$ are given for Hilbert spaces up to  dimension five.
We follow the notation and units of reference \cite{Carbone2002Quantum}, but find it crucial to apply a change  of phase $\Psi^{(k)}_{\ell}= (-i)^{\ell}\Phi^{(k)}_{\ell}$ to obtain a real, finite--difference Schrödinger--like equation
\begin{gather}\label{RR1}
\alpha_{\, \ell+1}\;\Phi_{\, \ell+1}^{(k)}
\,+\, \,\alpha_{\, \ell}\;\Phi_{\, \ell-1}^{(k)}=k \,\Phi_{\, \ell}^{(k)}\,.
\end{gather}
The $\Phi_{\, \ell}^{k}$ are the eigenfunctions of the volume operator 
expanded in the $\bf{J}_{12}=\bf{J}_{1}+\bf{J}_{2}$ 
basis\footnote{ \label{Foot:J23}
The treatment would be analogous had we chosen the $\bf{J}_{23}=\bf{J}_2+\bf{J}_3$ basis, with $\tilde{\ell}=j_{23 }.$
This construction  relies on  properties of the quadratic operator algebra generated by $(\bf{J}_{12})^2$, $(\bf{J}_{23})^2$ and $K$. Once chosen the eigenbasis of $(\bf{J}_{12})^2$, the other two operators are also tridiagonal and have the form of Eq. \eqref{RR1}.} 
The matrix elements $\alpha_\ell$ in \eqref{RR1} are given in terms of geometric quantities, namely
\begin{gather}\label{alphal}
\alpha_{\ell}=\frac{F(\ell;j_{1}+1/2;j_2+1/2)F(\ell;j_3+1/2;j_4+1/2)}{\sqrt{(2\ell+1)(2\ell-1)}},
\end{gather}
where  $F(A,B,C))  =\tfrac{1}{4} [(A+B+C)  
\ (-A+B+C)\ (A-B+C)\ (A+B-C)] ^{\tfrac{1}{2}}$ is the Archimedes' (``Heron's'') formula for the  area of a triangle with side lengths $A,B$ and $C$. Thus   $\alpha_\ell$ is  proportional to the product of the areas of  the two triangles sharing the side of length $\ell$ and forming a 
quadrilateral of sides $j_1+\tfrac{1}{2},\, j_2+\tfrac{1}{2},\, j_3+\tfrac{1}{2}$ and $j_4+\tfrac{1}{2}$, the parameters entering in Eq.  (\ref{alphal})\footnote{In Loop Quantum Gravity $j_i$ labels eigenvalues of the area operator %according to $A_i= 
$8\pi\,L_p\,\gamma \sqrt{(j_i(j_i+1)}$, where $L_p$ is the Planck length and $\gamma$ is the Immirzi parameter (see \textit{e.g.} \cite{Rovelli2007QG,HalBianchi2011}). 
}.
Note that the latter physically correspond to four quantum numbers associated to the quantum angular momenta $\bf{J}_1 ,\bf{J}_2 , \bf{J}_3$ and $\bf{J}_4 $. They appear to be all on the same footing, indicating that the volume operator can be thought of as acting  democratically on either a composite system of four objects with vanishing total angular momentum, or a system of three objects with total angular momentum $\mathbf{J}_4$.
%
%%%%%%%%%%%%%%%%%%%%%%%%%%%%%%%%
%% FIGURE QUADRILATERAL FIG 1 %%
%%%%%%%%%%%%%%%%%%%%%%%%%%%%%%%%
%
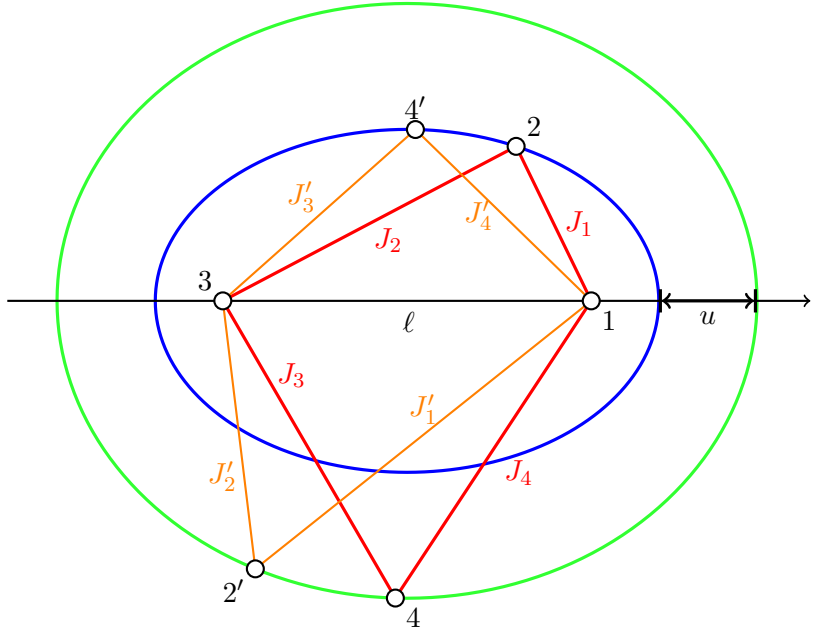
\begin{figure}[H,floatfix] %[htp]
\centering
\input{Fig1-EllipsesFlat.picture}

\caption{A quadrilateral and its Regge ``conjugate'' illustrating 
the elementary spin network representation of the symmetric coupling scheme: each 
quadrilateral  is dissected into two triangles sharing, as a common side, the diagonal $ \ell $.   The other sides are of  length $J_i=j_i +1/2$ (and $J'_i=j’_i+1/2$); $\ell$,  which is the discrete variable in Eq. \eqref{RR1}, is shown as the  distance between foci $1$ and $3$ of the confocal ellipses where the vertices of the quadrilaterals lie. 
The two sets of four side lengths of the Regge conjugate quadrilaterals are obtained by reflection with respect to the common semiperimeter $s$ (Eq. \eqref{eq:semiper}).  This relationship can be interpreted either as concerted stretchings and shortenings by the parameter  $r=  (j_1-j_2+j_3-j_4)/2$ introduced in \cite{Bitencourt2012Exact}, or by $v= (j_1-j_2-j_3+j_4)/2$ occurring in the projective interpretation of Robinson \cite{Robinson1970Group}. Shown is also the difference between the semimajor axes of the two ellipses, $ u=  (j_1+j_2-j_3-j_4)/2$.  Signs are decided according to the choice of primed and unprimed quadrilaterals. Also, $u$ and $v$ would exchange their roles had we chosen the other diagonal $\tilde{\ell}$ as the  variable $\ell$ (see note \ref{Foot:J23}).  In  
Eq. \eqref{eq:DefW} the orthogonal nature of this set of transformations is exhibited explicitly by the matrix  $W$. The passage to the Regge conjugate configuration $(s,-u,-r.-v)$ is revealed as a quaternionic conjugation, motivating our nomenclature.
}\label{fig:Quadrilateral}
\end{figure}
%%%%%%%%%%%%%%%%%%%%%%%%%%%%%%%%%%%%%%%%%
%% Endfigure
%%%%%%%%%%%%%%%%%%%%%%%%%%%%%%%%%%%%%%%%%%%

The similarities between the discrete Schrödinger equation and the three-term recursion occurring for the $6j$ symbol motivated the authors of  \cite{Carbone2002Quantum} to carry out the analysis  of its semiclassical behavior along the lines of  \cite{Ponzano1968Semiclassical,Schulten1975Semiclassical} . Another  similarity with the case of the $6j$ symbol  appears concerning  the range of $\ell$. As noted in \cite{Aquilanti2012Semiclassical},  the  basic requirement that the four vectors form a (not necessarily planar) quadrilateral leads to identify the range of $\ell$ with
\begin{gather}\label{DimK}
D\,=\,2\cdot\min\left(j_{1},j_{2},j_{3},j_{4},j'_{1},j'_{2},j'_{3},j'_{4}\right)+1 
\end{gather}
which is also the dimension of the Hilbert space where the volume operator acts. Here $j_i$ and $j'_i$ are conjugated by Regge  symmetry (Fig. \ref{fig:Quadrilateral}), \textit{i.e.} connected by $j'_i=s-j_i$, where
\begin{gather}\label{eq:semiper}
 s=\left(j_1+j_2+j_3+j_4\right)/2=\left(j'_1+j'_2+j'_3+j'_4\right)/2
\end{gather}
is the semiperimeter, common to both quadrilaterals. The map between primed and unprimed $j$s is  given by the symmetric $O(4)$ transformation in
% LARGE
\begin{gather}\label{eq:defR}
\frac{1}{2} 
\begin{pmatrix}
-1 & 1 & 1 & 1\\
1 & -1 & 1 & 1\\
1 & 1 & -1 & 1\\
1 & 1 & 1 & -1
\end{pmatrix}
\begin{pmatrix}
j_{1}\\
j_{2}\\
j_{3}\\
j_{4}
\end{pmatrix}
=
\begin{pmatrix}
j'_{1}\\
j'_{2}\\
j'_{3}\\
j'_{4}
\end{pmatrix},
\end{gather}
%
%% SMALL
%\begin{gather}\label{eq:defR}
%\tfrac{1}{2} 
%\left(
%\begin{smallmatrix}-1 & 1 & 1 & 1\\
%1 & -1 & 1 & 1\\
%1 & 1 & -1 & 1\\
%1 & 1 & 1 & -1
%\end{smallmatrix}
%\right)
%\left(
%\begin{smallmatrix}
%j_{1}\\
%j_{2}\\
%j_{3}\\
%j_{4}
%\end{smallmatrix}
%\right)
%=\left(
%\begin{smallmatrix}
%j'_{1}\\
%j'_{2}\\
%j'_{3}\\
%j'_{4}
%\end{smallmatrix}\right),
%\end{gather}
%
 denoted  $R$ in the following.
This is a striking manifestation of the relevance of the Regge symmetry in the present analysis: indeed it can be checked that the volume operator is  invariant under such  symmetry and therefore its spectrum and  eigenfunctions are invariant too. 
The Regge symmetry shows up also to be  important to assist in determining ranges of $\ell$ and of the $j$s. From Eq. \eqref{DimK} one can  decide to work with unprimed quantities, label as $j_1$ the minimum of the eight entries and ordering $j_2$ and $j_4$ according to $j_1 \leq j_2 \leq j_4$; then $j_2-j_1=\ell_m\leq \ell\leq j_1+j_2+1=\ell_M$ and $j_4 - j_2 +j_1 \leq j_3 \leq j_4+j_2-j_1$. The lower and upper limits in $j_3$ correspond respectively to $u=0$ and $r=0$, \textit{i.e.} cases when the two Regge conjugate quadrilaterals are coincident.

%\emph{Hamiltonian dynamics -} 
\section{Hamiltonian dynamics} 
\begin{figure} %[htp]
\centering	
\input{Fig2-Ellipses3D.picture}

\caption{The two quadrilaterals of Fig. \ref{fig:Quadrilateral}, looked at as a mechanical system, evolve creasing the pairs of triangles in which are dissected along $\ell$, according to a torsion mode corresponding to the same  dihedral angle $\frac{\pi}{2}+\varphi$ in both cases. Adding the edges $\overline{24}$ and $\overline{2'4'}$ two tetrahedra having the same volume can be visualized. In fact, their volume is proportional to  $H$ of Eq. \eqref{eq:ClHamilt} which is the product of the areas of two triangles  divided by the length of the hinging edge times the sine of the dihedral angle. Thus classically the volume is an energy function which is a constant of motion along the classical trajectories which are solutions of the Hamilton equations $ \frac{d\ell}{dt}=\frac{\partial H}{\partial \varphi}$; $ \frac{d \varphi}{dt}=-\frac{\partial H}{\partial \ell}$. Indeed, edges $\overline{24}$ and $\overline{2'4'}$ would have the same length $\tilde{\ell}=j_{23}$ had we  chosen to expand the volume operator in the basis of $\mathbf{J}_{23}=\bf{J}_2+\bf{J}_3$
(note \ref{Foot:J23}): two different confocal ellipses would describe the system and the vertices $2,4$ would coincide with  $2',4'$ as the foci of the new ellipses. On the other hand,  vertices $1$ and $3$ would split to give $1'$ and $3'$, say, lying on the new ellipses and  belonging either to a quadrilateral or to its conjugate. 
}
\label{Ellissi}
\end{figure}
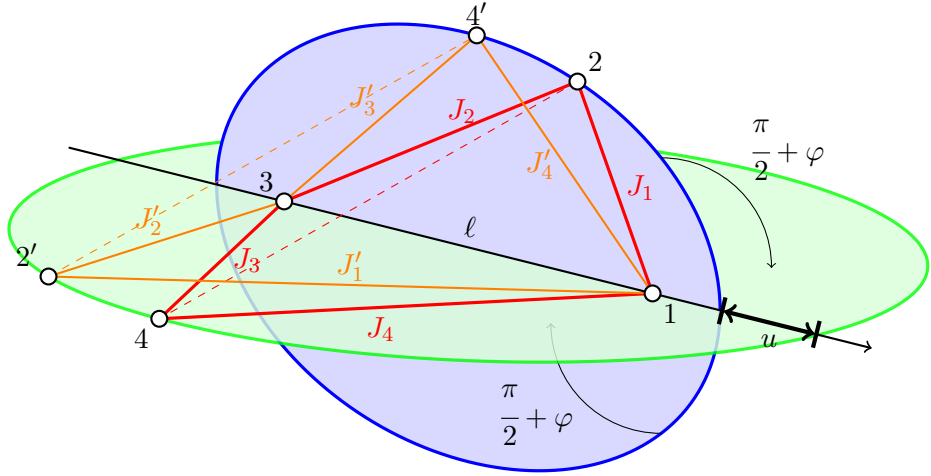

Transparent techniques are available  to study the semiclassical behavior of   difference equations of  type \eqref{RR1} (see. \textit{e.g.} \cite{Braun1993Discrete,Neville1971Technique} and references therein).
The  Hamiltonian operator for the discrete Schrödinger equation \eqref{RR1} can be written, in terms of the shift operator $ e^{\pm i \varphi} \Phi^{(k)}_{\ell}=\Phi^{(k)}_{\ell\pm 1}  $,
\begin{gather}\label{eq:QHamilt}
\hat{H}=\left(\alpha_{\ell}e^{-i\,\varphi}+\alpha_{\ell+1}
e^{i\,\varphi}\right) \quad \mbox{ with }\quad \varphi\,=\,-i\,\frac{\partial}{\partial\ \ell}
\end{gather}
representing the variable canonically conjugate  to $\ell$. The two-dimensional phase space $(\ell,\varphi)$  supports the  corresponding  classical Hamiltonian function  given by 
\begin{gather}\label{eq:ClHamilt}
H\,=\,2\,\alpha_{\ell+\frac{1}{2}}\cos\varphi,
\end{gather}
as illustrated in Fig. \ref{Ellissi} for the two Regge conjugate quadrilaterals of Fig. \ref{fig:Quadrilateral}, now allowed to fold along $\ell$ with $ \varphi $   perceived as a torsion angle\footnote{\label{Foot:ell}
Similarly the dependence on $ \ell $ can be appreciated  as a concerted bending mode, for example writing the product of the areas in the numerator of  Eq. \eqref{RR1} as $(J_1 J_2 J_3 J_4 \sin \theta_1 \sin \theta_3)/4$,  where $\theta_1$ and $\theta_3$ are internal angles in $1$ and $3$, respectively.% \sout{(see Figs. \ref{fig:Quadrilateral} and \ref{Ellissi})}. 
}.

\begin{figure} %[htp]
\centering
\includegraphics[scale=0.38]{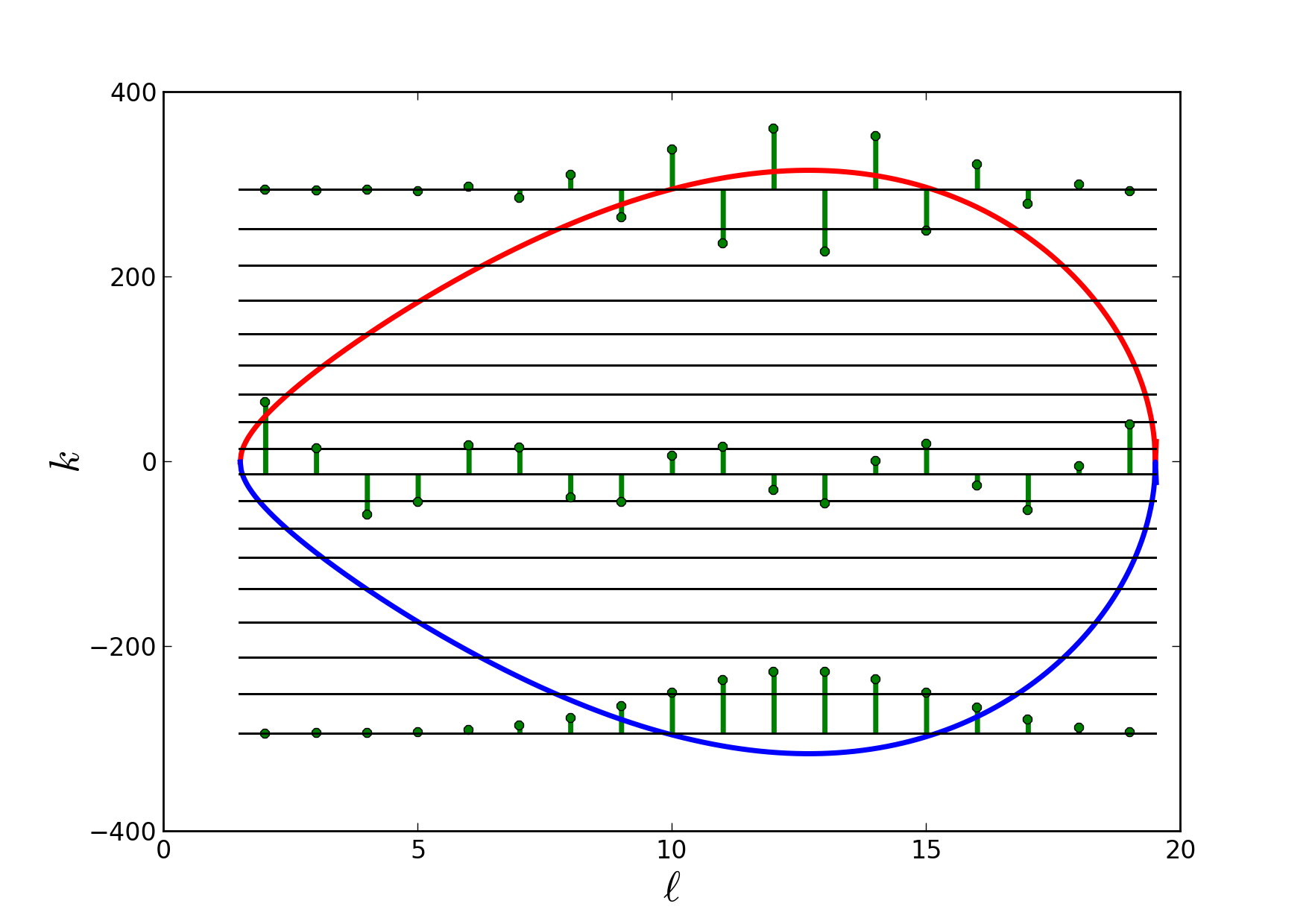}%
\includegraphics[scale=0.38]{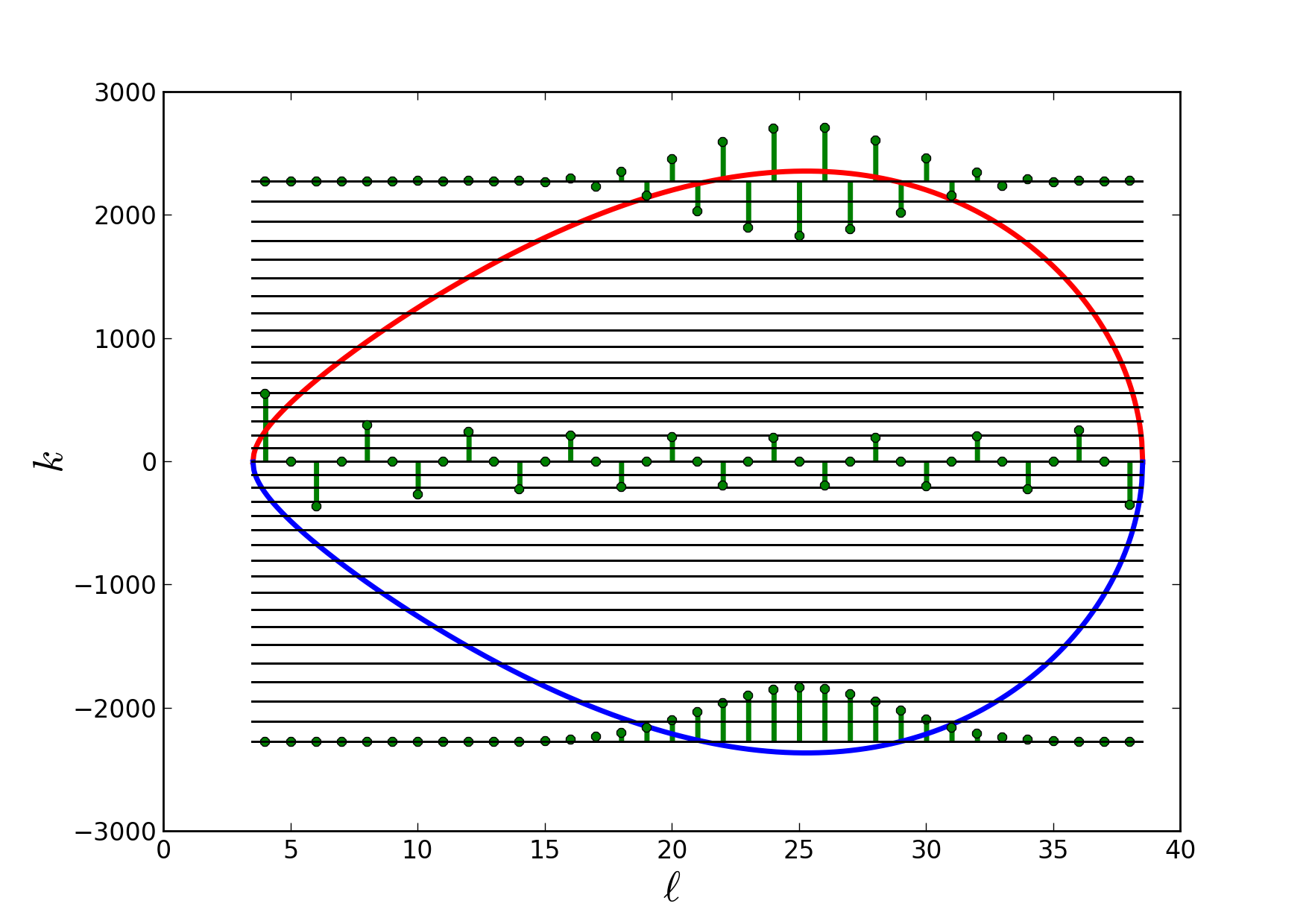} 

\caption{Two examples of spectra of the volume operator: the horizontal lines represent the eigenvalue $k$, the  curves are the caustics (the turning points of the semiclassical analysis), which limit  the classically allowed region (in red $U_{\ell}^+$, in blue $U_{\ell}^-$ Eq. \eqref{eq:DefPotentials}).
As can be seen, the eigenvalues are symmetrically distributed with respect to the $k=0$ (which is an eigenvalue if $D$ is odd). %(eigenvalues are not uniformly distributed)
In green the stick graph of three of the eigenfunctions (unnormalized). 
%Upper:
Left:
parameters $j_1,j_2,j_3,j_4$=8.5, 10.5, 13.5, 14.5
or $ s,u,r,v$=23.5, -4.5, 1.5, 0.5.
%Lower:
Right:
all four parameters are doubled.
The extrema  of $U^+_{\ell} $ and $U^{-}_{\ell}$ bracket the spectrum, which can be  well understood analytically and confirmed by extensive numerical  checks. 
The characteristic features of $U^+_{\ell} $ and $U^{-}_{\ell}$ can be compared to those for the caustics for the $6j$ symbol \cite{Bitencourt2012Exact}.
}
\label{Fig:Spectrum}
\end{figure}

\begin{figure} %[htp]
\centering
\includegraphics[scale=0.40]{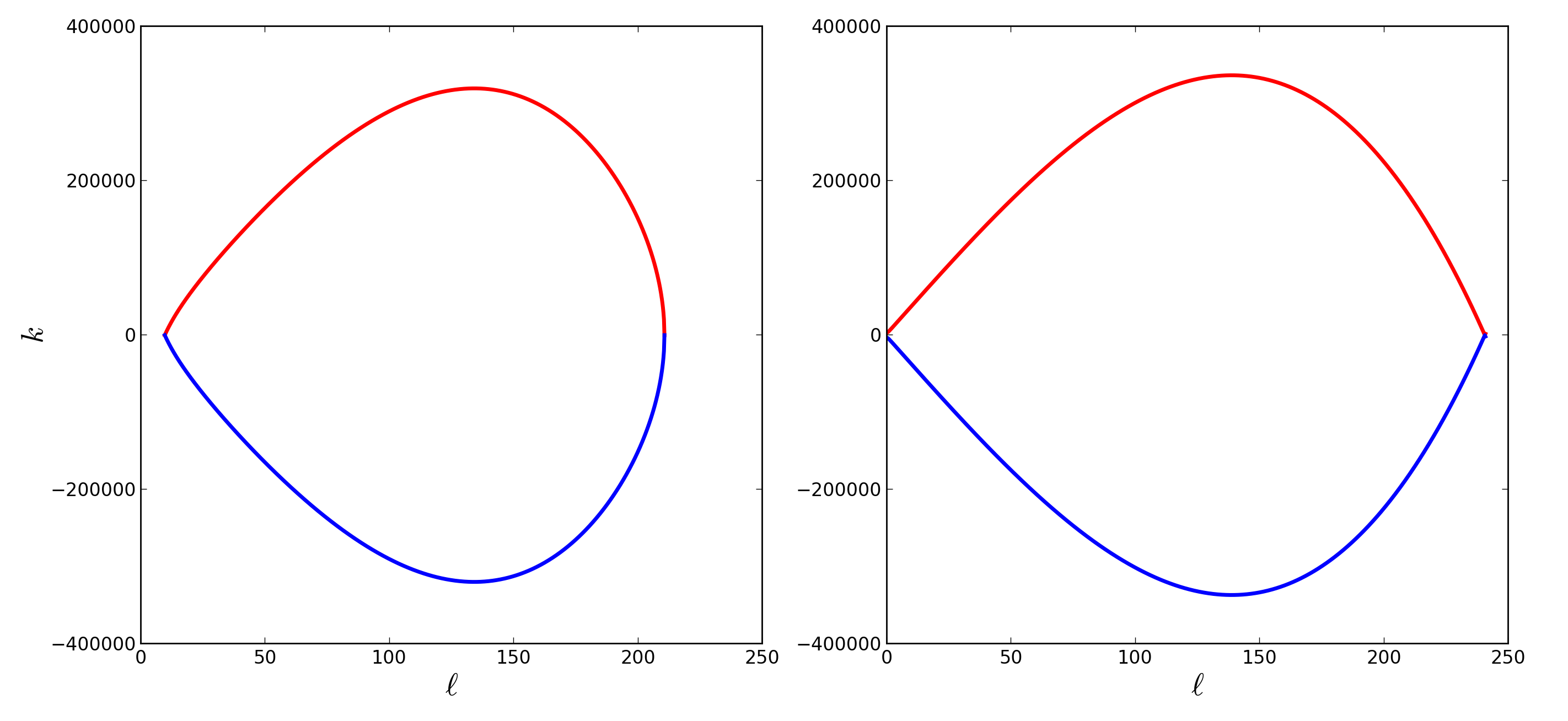}

\caption{
Potential functions $U^+$ and $U^-$ (Eq. \eqref{eq:DefPotentials}) are shown for two cases where the conjugated tetrahedra coincide. 
The cases occur when  (\emph{i}) -  $r=0$ (\textit{i.e.} $j_1+j_3=j_2+j_4$) (\textit{ii}) - $u=0$ ($j_1+j_2=j_3+j_4$). (\textit{iii}) $v=0$ ($j_1+j_4=j_2+j_3$). From the viewpoint of Fig. \ref{Ellissi}, case (\emph{i}) would correspond to a ``tangential'' quadrilateral, while (\textit{ii}) and (\textit{iii}) to ``ex-tangential'' ones.
In the left panel, case where only one of the $r,u,v$ variables
is zero ($j_1,j_2,j_3,j_4$=$100.0, 110.0, 130.0, 140.0$, $v=0$), while on the right all three of them are zero. The latter is the case for
an equilateral quadrilateral $j_1 = j_2 = j_3 = j_4 = 120.0 $ (compare to \cite{Bitencourt2012Exact} for the analogous cases which appear in the discussion of  the $6$j symbol).}\label{Fig:Caustics}
\begin{comment}
Figure 3:
V.O. with $j_1,j_2,j_3,j$=[100.0, 110.0, 130.0, 140.0]
$ \Rightarrow $$ s,u,r,v$=[240.0, -30.0, 10.0, 0.0] 
or $j_1^,,j_2^,,j_3^,,j$=[140.0, 130.0, 110.0, 100.0]
Figure 4:
V.O. with $j_1,j_2,j_3,j$=[120.0, 120.0, 120.0, 120.0]
$ \Rightarrow $$ s,u,r,v$=[240.0, 0.0, 0.0, 0.0] 
or $j_1^,,j_2^,,j_3^,,j$=[120.0, 120.0, 120.0, 120.0] 
\end{comment}
%
\end{figure}

The classical regime occurs when quantum numbers $j$s are large and $\ell$ can be considered as a continuous variable.
This limit for $\alpha_{\ell}$  permits us to draw the closed curves  in the $(\ell,k)$ plane of Fig. \ref{Fig:Spectrum} and Fig. \ref{Fig:Caustics}, obtained when $\varphi = 0$ or $\pi$ in Eq. \eqref{eq:ClHamilt}. These curves have the physical meaning of torsional-like potential functions
\begin{gather}\label{eq:DefPotentials}
U_{\ell}^+ =  - U_{\ell}^-=2\alpha_{\ell},
\end{gather}
 viewing the quadrilaterals in Figs. \ref{fig:Quadrilateral} and \ref{Ellissi} as mechanical systems. 
Noteworthy in Fig. \ref{Fig:Spectrum} is the further symmetry  along the $k = 0$ line ($\varphi = \pi /2$), missing \textit{e.g.} in the otherwise similar case of   $6j$ symbol  \cite{Bitencourt2012Exact}, where ``caustic curves'' are studied in a square of the $ \left( j_{12},j_{23} \right) $ plane (the ``screen''). Here the perfect duality between $j_{12}$ and $j_{23}$ is lost, and  the symmetry along the $k=0$ line appears in the potential functions as the continuous counterpart of the known fact that eigenvalues of the volume operator come in pairs, $k$ and $-k$  (plus  possibly the $k=0$ eigenvalue when $D$ is odd). Another manifestation of this symmetry links the eigenfunctions: it is easy to show from Eq. \eqref{RR1} that $\Phi^{(k)}_{\ell}=\left(-1 \right) ^{\ell} \Phi^{(-k)}_{\ell} $, and this appears in Fig. \ref{Fig:Spectrum} as the striking  alternating features in the positive part of the spectrum\footnote{The ``mirror'' symmetry of the $6j$ enlightened in \cite{Bitencourt2012Exact}  applies here too. In particular, allowing negative values of $\ell$ would permit infinite  replicas of Figs. \ref{Fig:Spectrum} and  \ref{Fig:Caustics} on both sides of the $\ell$ range.
}. 

The phenomenology of caustics presented in \cite{Bitencourt2012Exact} can be interestingly translated to this case, but taking into account also  such additional symmetries. 
Remarkably Regge symmetry is a key to the classification of the Lissajous-type  of potential functions given in  Eq. \eqref{eq:DefPotentials}. 
This includes also the limits where some quantities are large, which in the $6j$ case lead to the $3j$’s \cite{Bitencourt2012Exact}. In the present case this limiting procedure can be shown to lead to the cylindrical or planar spin networks discussed by Neville \cite{Neville2006Volume1,*Neville2006Volume2} for unpolarized and polarized gravitational waves.

%
%%%%%%%%%%%%%%%%%%%%%%%%%%%%%%%%%%%%%%%%%%%%%%%%%%%%%%
%% FIGURE ELIIPSE 3D Fig.2
%%%%%%%%%%%%%%%%%%%%%%%%%%%%%%%%%%%%%%%%%%%%%%%%%%%%%%%
%%%%%%%%%%%%%%%%%%%%%%%%%%%%%%%%%%%%%%%%%%%%%%%%%%%%%%%%
% END FIgurE
%%%%%%%%%%%%%%%%%%%%%%%%%%%%%%%%%%%%%%%%%%%%%%%%%%%%%%%

%\emph{Discrete orthogonal polynomials - }  
\section{Discrete orthogonal polynomials }\label{Sec:DOP}
The preceding considerations 
apply to the interesting issue of extracting the polynomial components out of wavefunctions and to this aim the defining three-term recursion in Eq. \eqref{RR1} is sufficient according to Favard theorem \cite{Atkinson196402}. Actually, these polynomials  can be  obtained from the secular equation once eigenvalues are  calculated. However, instead that directly from Eq. \eqref{RR1}, we find it much more insightful  (\emph{i}) - to eliminate the square roots to give an unsymmetrical three-term recursion with polynomial coefficients; this leads at each step to a polynomial proportional to $\Phi^{(k)}_{\ell}$ within a normalizing factor and a phase convention;
(\emph{ii}) - to impose Regge invariance as a guideline to obtain an  illuminating  geometrical interpretation, specifically that  of two triangles forming a tetrahedron, the two faces being hinged in a common side, and having respectively $s$ and $u$, or $r$ and $v$, as the other sides; 
(\emph{iii}) - to show that the  requirement of polynomial coefficients highlights the role of the new variables $s,u,r,v$,  introduced   on a purely  geometrical ground in Fig. \ref{fig:Quadrilateral}.
After some algebra, we obtain  from Eq. \eqref{RR1} the following  unsymmetrical three-term relation, which is manifestly Regge invariant and has polynomial coefficients
%\begin{multline}\label{eq:3TRRasym}
\begin{gather}\label{eq:3TRRasym}
\left(2\ell+1\right)  F^{2}\left(s,u,\ell-1\right) \, p_{\ell-1}^{\left(k\right)}
\,+\, \left(2\ell-1\right)F^{2}\left(r,v,\ell+1\right)\,p_{\ell+1}^{\left(k\right)}\,%=\\
=\,k\,\left(4\ell^{2}-1\right)p_{\ell}^{\left(k\right)}.
%\end{multline}
\end{gather}
The relation between $p^{(k)}_{\ell}$ and the $\Phi^{(k)}_{\ell}$ of Eq. \eqref{RR1} is given by 
\begin{equation}\label{Eq:2TRR}
p_{\ell}^{(k)}=N_{\ell}\Phi_{\ell}^{(k)} \mbox{ and } N_{\ell-1}=\frac{F\left(s,u,\ell-1\right)}{F\left(r,v,\ell\right)}N_{\ell},
\end{equation}
a two-term relation which  can be solved in closed form, see \textit{e.g.} \cite{QUA:QUA10497}. %,QUA:QUA10508}.
Normalization can be chosen by setting  boundary conditions (\textit{e.g.}  $ p^{(k)}_{\ell_{m}}=1,\ p^{(k)}_{\ell_{m-1}}=0 $).

%%%%%%%%%%%%%%%%%%%%%%%%%%%
%% FIGURE Spectrum Fig.3
%%%%%%%%%%%%%%%%%%%%%%%%%%%

%%%%%%%%%%%%%%%
%% End Spectrum
%%%%%%%%%%%%%%%

%%%%%%%%%%%%%%%%%%%%%%%%%%5
%% FIGURE Caustics Fig.4
%%%%%%%%%%%%%%%%%%%%%%%%%%%

%%%%%%%%%%%%%%%
%% End Caustics
%%%%%%%%%%%%%%%

%\emph{Concluding and further remarks-}  
\section{Concluding remarks and outlook}  
In retrospect, we have introduced a linear transformation $R$ which maps the two Regge conjugated quadrilaterals, Eq. \eqref{eq:defR}, and another  transformation $W$ which defines  the new variables (Fig. \ref{fig:Quadrilateral})
\begin{gather}\label{eq:DefW}
\frac{1}{2}
\begin{pmatrix}
	1 & 1 & 1 & 1\\
	1 & 1 & -1 & -1\\
	1 & -1 & -1 & 1\\
	1 & -1 & 1 & -1
\end{pmatrix}
\begin{pmatrix}
j_{1}\\
j_{2}\\
j_{3}\\
j_{4}
\end{pmatrix}
= 
\begin{pmatrix}
s\\
u\\
v\\
r
\end{pmatrix}.
\end{gather}
%%SMALL
%\begin{gather}\label{eq:DefW}
%\tfrac{1}{2}
%\left(
%\begin{smallmatrix}
%	1 & 1 & 1 & 1\\
%	1 & 1 & -1 & -1\\
%	1 & -1 & -1 & 1\\
%	1 & -1 & 1 & -1
%\end{smallmatrix}
%\right)
%\left(\begin{smallmatrix}
%j_{1}\\
%j_{2}\\
%j_{3}\\
%j_{4}
%\end{smallmatrix}\right)= \left(\begin{smallmatrix}
%s\\
%u\\
%v\\
%r
%\end{smallmatrix}\right).
%\end{gather}
The matrix $W$ is recognized as the famous one which provides  atomic ``hybrids'' of tetrahedral symmetry by combining one $s$ and three $p$ hydrogenoid orbitals   \cite{Pauling1931}.
Also, the  $(s,u,r,v)$ parametrization is reminiscent of   kinematic rotations \cite{PhysRevA.58.3718} %,QUA:QUA22763}
 of the quadrilaterals whose edges are interpreted as distances among four equal mass bodies.

Summarizing, the hidden Regge symmetry acts on  the  new variables  in an interesting way
\begin{gather}\label{Eq:CommDiag}
\xymatrix @R+2.0pc {
\ar@{}[dr]|{}
(j_{1},j_{2},j_{3},j_{4})\  \ar@<0.5ex>[d]\ar@<0.5ex>[r]^{R} &  \ (j'_{1},j'_{2},j'_{3},j'_{4})\ar@<0.5ex>[d]^{W}\ar@<0.5ex>[l]\\
\ (s,u,r,v)\ar@<0.5ex>[r]^{Q} \ar@<0.5ex>[u]^{{\huge W}}   &  \    (s,u',r'v')  \ar@<0.5ex>[u]\ar@<0.5ex>[l]
} 
\end{gather}
where $WRW=Q=\hbox{diag}(1,-1,-1,-1)$. Therefore  the transformation $W$ and the introduction of new variables permit to associate a quaternion and its conjugate to the two quadrilaterals twinned  by Regge symmetry. Implications of this remark in the mathematical context of dynamical algebras will be addressed elsewhere.

Regarding the quadrilaterals in Figs. \ref{fig:Quadrilateral} and \ref{Ellissi} as mechanical devices, note that the $u$, $v$ and $r$ coordinates are interestingly analogous to parameters occurring in the Grashof classification of four-bar linkages \cite{HartenbergDenavit1964}, the elementary moving mechanism of engines. For example, the conditions for identify of Regge conjugates, namely that at least one of them be zero, are those for the full folding of the mechanism.

 The educated guess that  equivolume  Regge conjugated tetrahedra are scissor congruent  eluded a constructive proof so far \cite{Roberts1999Classical,*Mohanty2003Regge}: our construction (Fig. \ref{Ellissi}) works by creasing along a diagonal the two plane isoperimetric quadrilaterals of Fig. \ref{fig:Quadrilateral}, concertedly stretched by either  $r$ or $u$ or $v$. The quadrilaterals are not scissor congruent: each is dissected into two triangles, with congruency with respect to the product, not the sum, of their areas.

  The discrete orthogonal family of polynomials  of section \ref{Sec:DOP}  is not ``classical'', \textit{i.e.} does not belong to  the hypergeometric families of the Askey  schemes (see \cite{QUA:QUA22117} for relevance in applied quantum mechanics).  Suitable  generalizations  would be interesting to be developed: indeed, the  situation is similar to that encountered for the Mathieu, Ince, Lam\`e and Heun families occurring for the separation of variables in elliptic coordinates for the action of the Laplacian operator on compact manifolds, see e.g. \cite{QUA:QUA10497,MendezFragoso2012Ladder}.
 Limiting cases can be formulated accordingly, such as the cylindrical or planar spin networks \cite{Neville2006Volume1,*Neville2006Volume2}, or the $q$ extensions\footnote{The $q$ extensions can be conveniently based on the formulation of $\alpha_\ell$ in Eq. \eqref{alphal}  as a  product of two $6j$'s \cite{JeanMarcLevyLeblond1965Symmetrical} for which the $q$-extension is well established}.

This set of issues, besides the cases outlined here, appears to be relevant to special function theory, with particular reference to the development of orthogonal basis sets of interest in applied quantum mechanics, and specifically in atomic and molecular physics, and in quantum chemistry.

We are grateful to  Mauro Carfora, Hal Haggard and  Robert Littlejohn for useful discussions.

\bibliography{VolumeOperator}

\end{document}

%% file: Fig1-EllipsesFlat.picture.tex
% author: Dimitri Marinelli
% distributed under Creative Commons -  Attribution 3.0 Unported 
% http://creativecommons.org/licenses/by/3.0/
% partially based on http://www.texample.net/tikz/examples/dandelin-spheres/
% by Hugues Vermeiren

\begin{tikzpicture}[PerspFlatFig1]%s,font=\large]
	
	% Iniziamo dichiarando la lunghezza dei vari ogetti
	\def\lla{14}
	\def\llb{27}
	\def\llx{30}
	\def\llc{28}
	\def\lld{29}

	\def\fangle{0}	%angolo di rotazioe
	
\begin{comment}
	\def\lla{10}
	\def\llb{10}
	\def\llx{10}
	\def\llc{10}
	\def\lld{10}
\end{comment}	

	% Semiperimetro!
	\pgfmathparse{(\lla+\llb+\llc+\lld)*0.5}
	\let\ss\pgfmathresult
	
	% normalizzo!
	\pgfmathparse{\lla/\ss}
	\let\la\pgfmathresult
	
	\pgfmathparse{\llb/\ss}
	\let\lb\pgfmathresult

	\pgfmathparse{\llc/\ss}
	\let\lc\pgfmathresult
	
	\pgfmathparse{\lld/\ss}
	\let\ld\pgfmathresult
	
	\pgfmathparse{\llx/\ss}
	\let\lx\pgfmathresult

	\pgfmathparse{(\la+\lb+\lc+\ld)*0.5}
	\let\s\pgfmathresult

	% Duali
	\pgfmathparse{\s-\la}
	\let\lap\pgfmathresult
	
	\pgfmathparse{\s-\lb}
	\let\lbp\pgfmathresult

	\pgfmathparse{\s-\lc}
	\let\lcp\pgfmathresult
	
	\pgfmathparse{\s-\ld}
	\let\ldp\pgfmathresult

	% Ellipses Center
	%ricordo che sebbene il disegno sia fatto sul piano y,z
	% nella omenclatura x->y e y->z
	\pgfmathparse{\lx/2}
	\let\XcE\pgfmathresult
	
	\def\YcE{0}

	% Raggio maggiore 
	
	%upper ellipse:
	\pgfmathparse{(\la+\lb)}
	\let\EuR\pgfmathresult	
	%lower ellipse
	\pgfmathparse{(\lc+\ld)}
	\let\ElR\pgfmathresult

	%Raggio minore
	
	%e' dato da (EllipseMaxRadSup**2-x**2)**0.5
	%upper ellipse
	\pgfmathparse{sqrt((\EuR)^2-(\lx^2))}
	\let\Eur\pgfmathresult  

	%lower ellipse
	\pgfmathparse{sqrt((\ElR)^2-(\lx^2))}
	\let\Elr\pgfmathresult

	%TESTS:
	\pgfmathparse{yTriang(\la,\lb,\lx)}
	\let\testm\pgfmathresult
	
	%TESTS:
	\pgfmathparse{xTriang(\la,\lb,\lx)}
	\let\testmx\pgfmathresult
\begin{comment}
	% the base circle is the unit circle in plane Oxy
	\def\h{0.5}% Heigth of the ellipse center (on the axis of the cylinder)
	\def\a{35}% angle of the section plane with the horizontal
	\def\aa{35}% angle that defines position of generatrix PA--PB
	\pgfmathparse{\h/tan(\a)}

  \let\b\pgfmathresult
	\pgfmathparse{sqrt(1/cos(\a)/cos(\a)-1)}
  \let\c\pgfmathresult %Center Focus distance of the section ellipse.
	\pgfmathparse{\c/sin(\a)}
  \let\p\pgfmathresult % Position of Dandelin spheres centers
                       % on the Oz axis (\h +/- \p)
                       
\end{comment}
%	\coordinate (1) at (2,\b,0);

	\coordinate (1) at (0,\lx,0);
	\coordinate (3) at (0,0,0);

	\coordinate (O) at (0,0,0);

	\coordinate (2) at ({-yTriang(\la,\lb,\lx)*sin(0)},{xTriang(\la,\lb,\lx)},{yTriang(\la,\lb,\lx)*cos(0)});
	\coordinate (4) at ({yTriang(\ld,\lc,\lx)*sin(\fangle)},{xTriang(\ld,\lc,\lx)},{-yTriang(\ld,\lc,\lx)*cos(\fangle)});
	
%	\draw (0) to node [below]  {$\overline{31}=\ell$} (1) ;
%    \draw[black,|<->|, ultra thick] (0) to node [below] {$\overline{31}=\ell$} (1);
	% assi y,z
%	\draw[->] (O)--(0.6,0,0)node[below left]{};

%	\draw[->] (O)--(0,0,0.3)node[left]{};

	% Gemelli

	\coordinate (2p) at ({yTriang(\lap,\lbp,\lx)*sin(\fangle)},{xTriang(\lap,\lbp,\lx)},{-yTriang(\lap,\lbp,\lx)*cos(\fangle)});
	\coordinate (4p) at ({-yTriang(\ldp,\lcp,\lx)*sin(0)},{xTriang(\ldp,\lcp,\lx)},{yTriang(\ldp,\lcp,\lx)*cos(0)});

	\coordinate (Tt) at (0,1.5,0);

	% Ellissi

	\coordinate (SEM) at (0,{(\XcE+\EuR/2)},0);
	\coordinate (SEm) at (0,{\XcE+\ElR/2},0);

	% dihedral angle
	
	\coordinate (dAs) at ({yTriang(\la,\lb,\lx)*sin(0)},{xTriang(\la,\lb,\lx)},{-yTriang(\la,\lb,\lx)*cos(0)});
	\coordinate (dAe) at ({yTriang(\la,\lb,\lx)*sin(\fangle)},{xTriang(\la,\lb,\lx)},{-yTriang(\la,\lb,\lx)*cos(\fangle)});

	\def\ttt{40};
	\def\fangles{0};
		\coordinate (dAs2) at ({(\YcE+\Eur/2*sin(\ttt))*(-sin(\fangles))},{\XcE+\EuR/2*cos(\ttt)},{(\YcE+\Eur/2*sin(\ttt))*(cos(\fangles))});
		
	\coordinate (dAe2) at ({(\YcE+\Eur/2*sin(\ttt))*(-sin(\fangle))},{\XcE+\EuR/2*cos(\ttt)},{(\YcE+\Eur/2*sin(\ttt))*(cos(\fangle))});

	\def\tttb{-40};
	\coordinate (dAs3) at ({(\YcE+\Eur/2*sin(\tttb))*(-sin(\fangles))},{\XcE+\EuR/2*cos(\tttb)},{(\YcE+\Eur/2*sin(\tttb))*(cos(\fangles))});
		
	\coordinate (dAe3) at ({(\YcE+\Eur/2*sin(\tttb))*(-sin(\fangle))},{\XcE+\EuR/2*cos(\tttb)},{(\YcE+\Eur/2*sin(\tttb))*(cos(\fangle))});

	%
		
	% Meta' Ellisse minore di sotto:
	\fill[white!15,draw=blue,very thick,opacity=1]
     (0,{(\XcE-\EuR/2)},0) % elliptical section
		\foreach \t in {182,184,...,360}
			{--(0,{\XcE+\EuR/2*cos(\t)},{\YcE+\Eur/2*sin(\t)})};

	% Draw diehdral angle
%	\draw [->] (dAs) arc [x radius=1 , y radius=1  ,start angle=180, end angle=0] ;

%	\fill[MyPoints] (dAs2) circle (0.3pt)node[below right]{$s$};
	
%	\draw [black,->,out=180,in=-90,] (dAs) to  node [ left] {$\theta$} (dAe);

%IUnvisibile in quello piatto
%	\draw [black,->,out=180,in=-90,] (dAs3) to  node [ left] {$\theta'$} (dAe3);

	% Ellisse, Maggiore:
	\draw[white!15,draw=green,very thick,opacity=0.81]
     (0,{\XcE+\ElR/2},0) % elliptical section
		\foreach \t in {2,4,...,360}
			{--({(\YcE+\Elr/2*sin(\t))*(-sin(\fangle))},{\XcE+\ElR/2*cos(\t)},{(\YcE+\Elr/2*sin(\t))*(cos(\fangle))})}--cycle;

	% Meta' minore, di sopra:
	\fill[white!15,draw=blue,very thick,opacity=1]
     (0,{\XcE+\EuR/2},0) % elliptical section
		\foreach \t in {2,4,...,180}
			{--(0,{\XcE+\EuR/2*cos(\t)},{\YcE+\Eur/2*sin(\t)})};

	% dihedral behind
%	\draw [black,->,out=0,in=90,] (dAs2) to  node [ above right] {$\theta'$} (dAe2);
			
	% asse x		
	\draw[->,  thick] (0,{(-\ElR/2+\XcE)*1.3},0)-- node[below]{$\ell$} (0,{(\ElR/2+\XcE)*1.1},0) ;

	% Draw Quadrilaterals:
	\draw[red,very thick] (1) to   node [right]  {$J_1$} (2) 
						  (2) to node [below right,inner sep=1pt] {$J_2$} (3) 
						  (3) to node [right,near start] {$J_3$} (4)
						  (4) to node [below right] {$J_4$} (1)--cycle;

%	\draw[red,dashed] (2)--(4);
	
	\draw[orange, thick] (1) to   node [left]  {$J'_4$} (4p) 
						  (4p) to node [above left,inner sep=1pt] {$J_3'$} (3) 
						  (3) to node [above left,near end] {$J_2'$} (2p)
						  (2p) to node [above] {$J_1'$} (1)--cycle;

%	\draw[orange,dashed] (2p)--(4p);

	% Freccia u
	\draw[black,|<->|, very thick] (SEM) to node [below] {$u$} (SEm);
	
	% names
	\fill[MyPoints] (1) circle (0.4pt)node[below right]{$1$};
	\fill[MyPoints] (2) circle (0.4pt)node[above right]{$2$};
	\fill[MyPoints] (3) circle (0.4pt)node[above left]{$3$};
	\fill[MyPoints] (4) circle (0.4pt)node[below right]{$4$};

	%dual
	%\fill[MyPoints] (1) circle (0.4pt)node[below right]{};
	%\fill[MyPoints] (3) circle (0.4pt)node[above right]{};
	\fill[MyPoints] (2p) circle (0.4pt)node[below left]{$2'$};
	\fill[MyPoints] (4p) circle (0.4pt)node[above ]{$4'$};

	%assi:
	\coordinate (O) at (0,0,0);

\begin{comment}
	\coordinate (O) at (0,0,0);
	\draw[->] (O)--(1,0,0)node[below left]{x};
	\draw[->] (O)--(0,1,0)node[right]{y};
	\draw[->] (O)--(0,0,1)node[left]{z};

\fill[MyPoints] (Tt) circle (0.1pt)node[right]{\EuR =\la+\lb};
\fill[MyPoints] (Tt) circle (0.1pt)node[above]{\Eur =\la+\lb};
\fill[MyPoints] (Tt) circle (0.1pt)node[below]{\testmx,\testm};

\end{comment}
\end{tikzpicture}

%% file: Fig2-Ellipses3D.picture.tex
% author: Dimitri Marinelli
% distributed under Creative Commons -  Attribution 3.0 Unported 
% http://creativecommons.org/licenses/by/3.0/
% partially based on http://www.texample.net/tikz/examples/dandelin-spheres/
% by Hugues Vermeiren
\begin{tikzpicture}[Persp3DFig2]%s,font=\large]
	
	% Iniziamo dichiarando la lunghezza dei vari ogetti
	\def\lla{14}
	\def\llb{27}
	\def\llx{30}
	\def\llc{28}
	\def\lld{29}

	\def\fangle{90}	%agolo di rotazioe
	
\begin{comment}
	\def\lla{10}
	\def\llb{10}
	\def\llx{10}
	\def\llc{10}
	\def\lld{10}
\end{comment}	

	% Semiperimetro!
	\pgfmathparse{(\lla+\llb+\llc+\lld)*0.5}
	\let\ss\pgfmathresult
	
	% normalizzo!
	\pgfmathparse{\lla/\ss}
	\let\la\pgfmathresult
	
	\pgfmathparse{\llb/\ss}
	\let\lb\pgfmathresult

	\pgfmathparse{\llc/\ss}
	\let\lc\pgfmathresult
	
	\pgfmathparse{\lld/\ss}
	\let\ld\pgfmathresult
	
	\pgfmathparse{\llx/\ss}
	\let\lx\pgfmathresult

	\pgfmathparse{(\la+\lb+\lc+\ld)*0.5}
	\let\s\pgfmathresult

	% Duali
	\pgfmathparse{\s-\la}
	\let\lap\pgfmathresult
	
	\pgfmathparse{\s-\lb}
	\let\lbp\pgfmathresult

	\pgfmathparse{\s-\lc}
	\let\lcp\pgfmathresult
	
	\pgfmathparse{\s-\ld}
	\let\ldp\pgfmathresult

	% Ellipses Center
	%ricordo che sebbene il disegno sia fatto sul piano y,z
	% nella omenclatura x->y e y->z
	\pgfmathparse{\lx/2}
	\let\XcE\pgfmathresult
	
	\def\YcE{0}

	% Raggio maggiore 
	
	%upper ellipse:
	\pgfmathparse{(\la+\lb)}
	\let\EuR\pgfmathresult	
	%lower ellipse
	\pgfmathparse{(\lc+\ld)}
	\let\ElR\pgfmathresult

	%Raggio minore
	
	%e' dato da (EllipseMaxRadSup**2-x**2)**0.5
	%upper ellipse
	\pgfmathparse{sqrt((\EuR)^2-(\lx^2))}
	\let\Eur\pgfmathresult  

	%lower ellipse
	\pgfmathparse{sqrt((\ElR)^2-(\lx^2))}
	\let\Elr\pgfmathresult

	%TESTS:
	\pgfmathparse{yTriang(\la,\lb,\lx)}
	\let\testm\pgfmathresult
	
	%TESTS:
	\pgfmathparse{xTriang(\la,\lb,\lx)}
	\let\testmx\pgfmathresult
\begin{comment}
	% the base circle is the unit circle in plane Oxy
	\def\h{0.5}% Heigth of the ellipse center (on the axis of the cylinder)
	\def\a{35}% angle of the section plane with the horizontal
	\def\aa{35}% angle that defines position of generatrix PA--PB
	\pgfmathparse{\h/tan(\a)}

  \let\b\pgfmathresult
	\pgfmathparse{sqrt(1/cos(\a)/cos(\a)-1)}
  \let\c\pgfmathresult %Center Focus distance of the section ellipse.
	\pgfmathparse{\c/sin(\a)}
  \let\p\pgfmathresult % Position of Dandelin spheres centers
                       % on the Oz axis (\h +/- \p)
                       
\end{comment}
%	\coordinate (1) at (2,\b,0);

	\coordinate (1) at (0,\lx,0);
	\coordinate (3) at (0,0,0);

	\coordinate (O) at (0,0,0);

	\coordinate (2) at ({-yTriang(\la,\lb,\lx)*sin(0)},{xTriang(\la,\lb,\lx)},{yTriang(\la,\lb,\lx)*cos(0)});
	\coordinate (4) at ({yTriang(\ld,\lc,\lx)*sin(\fangle)},{xTriang(\ld,\lc,\lx)},{-yTriang(\ld,\lc,\lx)*cos(\fangle)});
	
%	\draw (0) to node [below]  {$\overline{31}=\ell$} (1) ;
%    \draw[black,|<->|, ultra thick] (0) to node [below] {$\overline{31}=\ell$} (1);
	% assi y,z
%	\draw[->] (O)--(0.6,0,0)node[below left]{};

%	\draw[->] (O)--(0,0,0.3)node[left]{};

	% Gemelli

	\coordinate (2p) at ({yTriang(\lap,\lbp,\lx)*sin(\fangle)},{xTriang(\lap,\lbp,\lx)},{-yTriang(\lap,\lbp,\lx)*cos(\fangle)});
	\coordinate (4p) at ({-yTriang(\ldp,\lcp,\lx)*sin(0)},{xTriang(\ldp,\lcp,\lx)},{yTriang(\ldp,\lcp,\lx)*cos(0)});

	\coordinate (Tt) at (0,1.5,0);

	% Ellissi

	\coordinate (SEM) at (0,{(\XcE+\EuR/2)},0);
	\coordinate (SEm) at (0,{\XcE+\ElR/2},0);

	% dihedral angle
	
	\coordinate (dAs) at ({yTriang(\la,\lb,\lx)*sin(0)},{xTriang(\la,\lb,\lx)},{-yTriang(\la,\lb,\lx)*cos(0)});
	\coordinate (dAe) at ({yTriang(\la,\lb,\lx)*sin(\fangle)},{xTriang(\la,\lb,\lx)},{-yTriang(\la,\lb,\lx)*cos(\fangle)});

	\def\ttt{40};
	\def\fangles{0};
		\coordinate (dAs2) at ({(\YcE+\Eur/2*sin(\ttt))*(-sin(\fangles))},{\XcE+\EuR/2*cos(\ttt)},{(\YcE+\Eur/2*sin(\ttt))*(cos(\fangles))});
		
	\coordinate (dAe2) at ({(\YcE+\Eur/2*sin(\ttt))*(-sin(\fangle))},{\XcE+\EuR/2*cos(\ttt)},{(\YcE+\Eur/2*sin(\ttt))*(cos(\fangle))});

	\def\tttb{-40};
	\coordinate (dAs3) at ({(\YcE+\Eur/2*sin(\tttb))*(-sin(\fangles))},{\XcE+\EuR/2*cos(\tttb)},{(\YcE+\Eur/2*sin(\tttb))*(cos(\fangles))});
		
	\coordinate (dAe3) at ({(\YcE+\Eur/2*sin(\tttb))*(-sin(\fangle))},{\XcE+\EuR/2*cos(\tttb)},{(\YcE+\Eur/2*sin(\tttb))*(cos(\fangle))});

	%
		
	% Meta' Ellisse minore di sotto:
	\fill[blue!15,draw=blue,very thick,opacity=1]
     (0,{(\XcE-\EuR/2)},0) % elliptical section
		\foreach \t in {182,184,...,360}
			{--(0,{\XcE+\EuR/2*cos(\t)},{\YcE+\Eur/2*sin(\t)})};

	% Draw diehdral angle
%	\draw [->] (dAs) arc [x radius=1 , y radius=1  ,start angle=180, end angle=0] ;

%	\fill[MyPoints] (dAs2) circle (0.3pt)node[below right]{$s$};
	
%	\draw [black,->,out=180,in=-90,] (dAs) to  node [ left] {$\theta$} (dAe);

	\draw [black,->,out=180,in=-90,] (dAs3) to  node [ left] {$\cfrac{\pi}{2}+\varphi$} (dAe3);

	% Ellisse, Maggiore:
	\fill[green!15,draw=green,very thick,opacity=0.81]
     (0,{\XcE+\ElR/2},0) % elliptical section
		\foreach \t in {2,4,...,360}
			{--({(\YcE+\Elr/2*sin(\t))*(-sin(\fangle))},{\XcE+\ElR/2*cos(\t)},{(\YcE+\Elr/2*sin(\t))*(cos(\fangle))})}--cycle;

	% Meta' minore, di sopra:
	\fill[blue!15,draw=blue,very thick,opacity=1]
     (0,{\XcE+\EuR/2},0) % elliptical section
		\foreach \t in {2,4,...,180}
			{--(0,{\XcE+\EuR/2*cos(\t)},{\YcE+\Eur/2*sin(\t)})};

	% dihedral behind
	\draw [black,->,out=0,in=90,] (dAs2) to  node [ above right] {$\cfrac{\pi}{2}+\varphi$} (dAe2);
			
	% asse x		
%	\draw[->,  thick] (0,{(-\ElR/2+\XcE)*1.3},0)--(0,{(\ElR/2+\XcE)*1.1},0) node[right]{$\overline{31} \equiv \ell$};
	\draw[->,  thick] (0,{(-\ElR/2+\XcE)*1.3},0)-- node[above]{$ \ell$} (0,{(\ElR/2+\XcE)*1.1},0);

	% Draw Quadrilaterals:
	\draw[red,very thick] (1) to   node [right]  {$J_1$} (2) 
						  (2) to node [above,inner sep=1.6pt,pos=0.395] {$J_2$} (3) 
						  (3) to node [right] {$J_3$} (4)
						  (4) to node [below left] {$J_4$} (1)--cycle;

	\draw[red,dashed] (2)--(4);
	
	\draw[orange, thick] (1) to   node [left]  {$J'_4$} (4p) 
						  (4p) to node [above left,inner sep=1pt] {$J_3'$} (3) 
						  (3) to node [above left,inner sep=1pt] {$J_2'$} (2p)
						  (2p) to node [above,inner sep=1pt] {$J_1'$} (1)--cycle;
	\draw[orange,dashed] (2p)--(4p);

	% Freccia u
	\draw[black,|<->|, ultra thick] (SEM) to node [below] {$u$} (SEm);
	
	% names
	\fill[MyPoints] (1) circle (0.3pt)node[below right]{$1$};
	\fill[MyPoints] (2) circle (0.3pt)node[above right]{$2$};
	\fill[MyPoints] (3) circle (0.3pt)node[above left]{$3$};
	\fill[MyPoints] (4) circle (0.3pt)node[below left]{$4$};

	%dual
	\fill[MyPoints] (2p) circle (0.3pt)node[above left]{$2'$};
	\fill[MyPoints] (4p) circle (0.3pt)node[above ]{$4'$};

	%assi:
	\coordinate (O) at (0,0,0);

\begin{comment}
	\coordinate (O) at (0,0,0);
	\draw[->] (O)--(1,0,0)node[below left]{x};
	\draw[->] (O)--(0,1,0)node[right]{y};
	\draw[->] (O)--(0,0,1)node[left]{z};

\fill[MyPoints] (Tt) circle (0.1pt)node[right]{\EuR =\la+\lb};
\fill[MyPoints] (Tt) circle (0.1pt)node[above]{\Eur =\la+\lb};
\fill[MyPoints] (Tt) circle (0.1pt)node[below]{\testmx,\testm};

\end{comment}
\end{tikzpicture}